\documentclass[11pt,a4paper]{amsart}
\usepackage{amsmath,amssymb}
\usepackage{amsfonts}
\usepackage{eucal}
\usepackage{amsthm}
\numberwithin{equation}{section}

\def\a{\alpha}

\def\c{\gamma}
\def\d{\delta}

\def\f{\varphi}
\def\g{\psi}
\def\h{\hbar}
\def\i{\mbox{\raisebox{.5ex}{$\chi$}}}

\def\l{\lambda}
\def\m{\mu}
\def\n{\nu}

\def\t{\tau}
\def\x{\xi}
\def\y{\eta}
\def\z{\zeta}

\newcommand{\Op}{\mathrm{Op}}

\def\re{\mathbb{R}}
\def\co{\mathbb{C}}
\def\ze{\mathbb{Z}}

\def\pa{\partial}

\renewcommand{\Im}{\mathrm{Im}}

\newcommand{\supp}{\mathrm{supp}}

\newcommand{\Ker}{\mathrm{Ker}}

\newcommand{\norm}[1]{\| #1 \|}

\newcommand{\bigpare}[1]{\bigl(#1\bigr)}
\newcommand{\biggpare}[1]{\biggl(#1\biggr)}
\newcommand{\Bigpare}[1]{\Bigl(#1\Bigr)}

\newcommand{\bigbra}[1]{\bigl\{#1\bigr\}}

\newcommand{\bigset}[2]{\bigl\{#1\bigm|#2\bigr\}}

\newcommand{\jap}[1]{\langle #1 \rangle}

\newcommand{\abs}[1]{| #1 |}
\newcommand{\bigabs}[1]{\bigl| #1 \bigr|}
\newcommand{\Bigabs}[1]{\Bigl| #1 \Bigr|}

\newcommand{\beq}{\begin{equation}}
\newcommand{\eeq}{\end{equation}}
\newcommand{\ba}{\begin{align}}
\newcommand{\ea}{\end{align}}

\newtheorem{thm}{Theorem}[section]
\newtheorem{lem}[thm]{Lemma}

\theoremstyle{definition}

\newtheorem{ass}{Assumption}

\theoremstyle{remark}
\newtheorem{rem}{Remark}

\newcommand{\WF}{\mathrm{W\!F}}

\title[Essential self-adjointness for Klein-Gordon type operators]%
{Essential self-adjointness for the Klein-Gordon type operators on asymptotically static spacetime}

\author{Shu Nakamura}
\address{Department of Mathematics,
Faculty of Sciences,
Gakushuin University,
1-5-1, Mejiro, Toshima, Tokyo
171-8588, Japan.}
\email{shu.nakamura@gakushuin.ac.jp} 
\author{Kouichi Taira}
\address{Department of Mathematical Sciences, Ritsumeikan University, 1-1-1 NojiHigashi, 
Kusatsu, 525-8577 Japan.}
\email{ktaira@fc.ritsumei.ac.jp }

\begin{document}
\maketitle

\begin{abstract}
Let $X=\re\times M$ be the spacetime, 
where $M$ is a closed manifold equipped with a Riemannian metric $g$, 
and we consider a symmetric Klein-Gordon type operator $P$ on $X$, which is 
asymptotically converges to $\pa_t^2-\triangle_g$ as $|t|\to\infty$, 
where $\triangle_g$ is the Laplace-Beltrami operator on $M$. 
We prove the essential self-adjointness of $P$ on $C_0^\infty(X)$.
The idea of the proof is closely related to a recent paper by the authors on the essential self-adjointness 
for Klein-Gordon operators on asymptotically flat spaces. 
\end{abstract}

\section{Introduction}\label{sec-introduction} 
Let $M$ be an $n$-dimensional closed Riemannian manifold with the metric $g$, and let
\[
X=\re\times M
\]
be our \textit{spacetime}. We consider a Klein-Gordon type 
operator corresponding to a Lorentz metric on $X$, 
which we denote by $P$. Let $p(t,x,\t,\x)$ be the symbol of $P$ on $T^*X$, where 
$(t,\t)\in T^*\re\cong\re\times\re$, $(x,\x)\in T^*M$. We suppose $p$ is nondegenerate quadratic form 
with respect $(\t,\x)$ for each $(t,x)$, and it has the form:
\[
p(t,x,\t,\x)= \t^2-q_0(x,\x)+q(t,x,\t,\x),
\]
where 
\[
q_0(x,\x)=\sum_{j,k} g^{ik}(x)\x_j\x_k
\]
is the standard cometric on $T^*M$. We note the mass term is irrelevant 
to the self-adjointness, and we assume the mass is zero without loss 
of generality. 
We suppose 

\begin{ass}\label{ass-main}
$q(t,x,\t,\x)$ is smooth and there is $\m>0$ such that for any multi-index 
$\a\in \ze_+^{n}$ and $k\in\ze_+$,  
\[
\bigabs{\pa_t^k \pa_x^\a q(t,x,\t,\x)}\leq 
C_{k,\a}\jap{t}^{-1-\m} (1+\t^2+|\x|^2).
\]
\end{ass}

We need a sort of nontrapping condition. 
Let $\exp(sH_p)$ be the Hamilton flow on $T^*X$ generated by the symbol $p$, and we denote
a geodesic by 
\[
(t(s),x(s),\t(s),\x(s))=\exp(sH_p)(t_0,x_0,\t_0,\x_0),
\]
where $(t_0,x_0,\t_0,\x_0)\in T^*X$ is the 
initial condition of the geodesic. A geodesic is called \textit{null}\/ if $p(t(s),x(s),\t(s),\x(s))=0$ 
(for some $s$, and hence for all $s$). 
Then we assume the following: 

\begin{ass}\label{ass-time-like}
Any null geodesic satisfies either $t(s)\to\pm\infty$ as $s\to\pm\infty$ or 
$t(s)\to\mp\infty$ as $s\to\pm\infty$. 
\end{ass}

\begin{thm}\label{thm-main}
Suppose Assumptions~\ref{ass-main} and \ref{ass-time-like}. Then $P$ is 
essentially self-adjoint on $C_0^\infty(X)$. 
\end{thm}

We note the result is stable under bounded perturbations, and hence we may 
add bounded potentials without changing the essential self-adjointness 
(see, e.g., \cite{RS2}). 

One motivation to show the essential self-adjointness for Klein-Gordon operators on spacetime 
is the construction of Feynman propagators on curved spacetime, which is crucial in 
the construction of quantum field theory. The Feynman propagator is 
formally constructed as a boundary value of the resolvent for the Klein-Gordon operator 
on the spacetime, but the self-adjointness of these Klein-Gordon operator had not been proved 
except for the stationary cases. See papers by Derezi\'{n}ski-Siemssen \cite{DeSi1,DeSi2,DeSi3} and 
G\'{e}rard-Wrochna \cite{GeWr1,GeWr2} about the background. 
The essential self-adjointness of Klein-Gordon operators was conjectured in these papers, 
and it was proved for asymptotically flat spacetime by Vasy \cite{Va} and Nakamura-Taira
\cite{NT21}. Recently the authors obtained a simplified proof of this result \cite{NT22}, 
and this paper employs closely related argument to solve the problem for asymptotically 
static spacetime, which has been also conjectured and open so far. For other motivations and related 
topics in scattering theory, we refer these papers \cite{Va,NT21,NT22} and references therein. 

We prove our main theorem in Section~\ref{sec-proof-of-themorem-main}.  
The argument generally follows that of \cite{NT22}, and actually somewhat simpler, 
though there are several significant differences, 
in particular, concerning the construction of escaping operator $B$. 
A technical lemma is proved in Appendix~\ref{app-microlocalize-proof}. 


\section{Proof of Theorem~\ref{thm-main}}
\label{sec-proof-of-themorem-main}

We use the following notations throuout this paper. 
Let $a$ be a symbol: $a(h;\cdot,\cdot)\in C^\infty(T^*X)$, 
$\f\in C_0^\infty(X)$ and $h>0$. Then we denote the Weyl quantization with the semiclassical 
parameter $h$ by $\Op_h(a)$ on $X$. Namely, 
\begin{multline*}
\Op_h(a)\f(t,x)\\ =(2\pi h)^{-n-1}\iint_{T^*X} e^{i(t\t+(x-y)\cdot\x)/\h}
a\bigpare{\tfrac{t+s}{2},\tfrac{x+y}{2},\t,\x} \f(s,y)dsdyd\t d\x
\end{multline*}
locally in $x\in M$, and globally in $t\in \re$. We emphasis that our pseudodifferential 
operator calculus is local in $x\in M$, but global in $t\in\re$. 
We also denote $\Op(\cdot)=\Op_1(\cdot)$, 
i.e., the non-semiclassical Weyl quantization. We refer, e.g., Dimassi-Sj\"ostrand \cite{DiSj},
or Zworski \cite{Zw} for the semiclassical pseudodifferential operator calculus. 

For a pair of symbols $a(t,x,\t,\x)$ and $b(t,x,\t,\x)$, we denote the Poisson bracket by 
\[
\{a,b\}=\frac{\pa a}{\pa \t}\frac{\pa b}{\pa t} +\frac{\pa a}{\pa \x}\cdot \frac{\pa b}{\pa x}
- \frac{\pa a}{\pa t}\frac{\pa b}{\pa \t} -\frac{\pa a}{\pa x}\cdot \frac{\pa b}{\pa \x}
=\frac{d}{ds}\exp(sH_a)b\big|_{s=0}.
\]
Let $Q_0=\Op(q_0)$. Since $Q_0$ is the Laplace-Beltrami operator, 
it is well-known that $Q_0$ is essentially self-adjoint 
on $C_0^\infty(M)$. We denote the unique self-adjoint extension by the same symbol $Q_0$. 
We may write
\[
P=D_t^2-Q_0+Q, \quad Q=\Op(q).
\]

We denote $\i_1(s)\in C^\infty(\re)$ such that $\i_1(s)=1$ for $s\leq -1$, $\supp[\i_1]\subset(-\infty,0]$, 
and $\i_1'(s)\leq 0$ for all $s\in\re$. We also denote 
$\i_2\in C_0^\infty(\re)$ such that $\i_2(s)=1$ for $s\in [-1,1]$, $\supp[\i_2]\subset [-2,2]$, 
and $s\i_2'(s)\leq 0$ for all $s\in\re$. 

\subsection{The first reduction}
\label{subsec-the-first-reduction}

By the basic criterion for the essential self-adjointness, it suffices to show 
$\Ker(P^*-z_\pm)=\{0\}$ for $z_{\pm}\in\co$, $\pm\Im z_\pm>0$ ( \cite{RS2} Theorem~X.1)
We note $(P^*-z)\g=0$ is equivalent to $(P-z)\g=0$ in the distribution sense, 
since the domain of $P$ is $\mathcal{D}=C_0^\infty(X)$. 
Thus we prove that if $(P-z_\pm)\g=0$ for $\g \in L^2(X)$ in the distribution sense  
then $\g=0$. Here is a simple but useful condition to show $\g=0$. 
We denote the Sobolev space of order $s$ on $X$ by $H^s(X)$. 

\begin{lem}\label{lem-first-reduction}
Let $\g\in L^2(X)$ such that $(P-z)\g=0$ with $z\in\co$. 
If, moreover, $\jap{t}^{-1}\g\in H^{1}(X)$, then $\g=0$.
\end{lem}

The proof is a simple commutator computation, and we omit it. See, e.g., Appendix~A 
of \cite{NT22}. We also note that actually the condition $\jap{t}^{-1/2}\g\in H^{1/2}(X)$ 
is sufficient, and this is used in \cite{Va} and \cite{NT21}, though our condition 
is sufficient for our purpose and the proof is slightly more elementary. 

In the following, we will show that if $(P-z)\g=0$ with $z\in\co$, $\Im z>0$ then 
$\jap{t}^\c\g\in H^N(X)$ with any $\c, N>0$. The case $\Im z<0$ is similar, 
and we mostly concentrate on the case $\Im z>0$.

\subsection{Remarks on microlocal regularities}
\label{subsec-remarks-on-microlocal-regularities}

We use the well-known semiclassical characterization of the microlocal regularities. 
Let $\g\in\mathcal{D}'(\re^n)$ and $(x_0,\x_0)\in\re^{2n}$ with $\x_0\neq 0$. 
Then $(x_0,\x_0)\notin \WF(\g)$ if there is $a\in C_0^\infty(\re^{2n})$ 
such that $a(x_0,\x_0)\neq 0$ and 
\[
\norm{\Op_h(a)\g}_{L^2}=\norm{a(x,hD_x)\g}_{L^2}=O(h^\infty), \quad 
\text{as }h\to 0, 
\]
where $\WF(\g)$ is the wave front set of $\g$. 
It is also easy to show, under the same notation, $\g$ is microlocally $H^s$ near $(x_0,\x_0)$ 
if $\norm{\Op_h(a)\g)}_{L^2}=O(h^{s'})$ as $h\to 0$ with some $s'>s$. 
This characterization works globally also. For example, suppose $a\in S^0_{1,0}$ and 
$a(x,\x)\geq c_0>0$ if  $\bigabs{|\x|-1}<\d$ with $\d>0$, and if $\norm{\Op_h(a)\g}_{L^2}=O(h^{s'})$ as $h\to 0$, 
then $\g\in H^s$ for $s<s'$. 

We now consider our spacetime model. We note that away from the characteristic set: 
\[
\mathrm{Char}(P)=\bigset{(t,x,\t,\x)}{p(t,x,\t,\x)=0},
\]
the operator $P$ is elliptic, and hence $(P-z)\g=0$ and $(t_0,x_0,\t_0,\x_0)\notin \mathrm{Char}(P)$ 
imply $(t_0,x_0,\t_0,\x_0)\notin\WF(\g)$. 
We note moreover that, on the conic set: 
\[
\bigset{(t,x,\t,\x)}{|p(t,x,\t,\x)|>\d(\t^2+q_0(x,\x))}, \quad \d>0,
\]
the operator $P$ is uniformly elliptic, and hence
$\g\in H^\infty$ there with suitable microlocal cut-off. 
Thus, in order to show the smoothness of $\g$, it suffices to study microlocal smoothness 
in a neighborhood of $\mathrm{Char}(P)$. 

We first consider the area $|t|\gg 0$. As $|t|\to\infty$, by Assumption~\ref{ass-main}, 
$p(x,\x)\sim \t^2-q_0(x,\x)$, and hence $\t^2\sim q_0(x,\x)$ on $\mathrm{Char}(P)$. 
On the other hand, when we study the regularities of $\g$ in a neighborhood of $(t,x,\t,\x)$
using the semiclassical method, 
we consider the area $\t^2+q_0(x,\x)\sim c_0 h^{-2}$ with some $c_0>0$, e.g., $c_0=2$. 
Now, if $\t^2\sim q_0(x,\x)$ and $\t^2+q_0(x,\x)\sim 2 h^{-2}$, then we conclude 
$\t^2\sim h^{-2}$ and $q_0(x,\x)\sim h^{-2}$. 
This informal argument suggests that it suffices to study the behavior of 
$\norm{\Op_h(a(t,x,\t,\x))\g}_{L^2}$ as $h\to 0$ in order to obtain regularities for $|t|\gg 0$, 
where $a$ is supported in a neighborhood of 
$\bigset{(t,x,\t,\x)}{\t^2=1,q_0(x,\x)=1}$. We actually use the following lemma: 

\begin{lem}\label{lem-microlocalize}
Let $\d>0$, and suppose $\g\in L^2$ with $(P-z)\g=0$. Then there is $T_0>0$ such that the following holds: 
Let 
\[
a(\d,T;t,x,\t,\x)= \i_2(|\t^2-1|/\d)\i_2(|q_0(x,\x)-1|/\d)\i_1(1-|t|/T). 
\]
If $\norm{\Op_h(a(\d,T))\g}_{L^2}=O(h^{s'})$ as $h\to 0$ with $T>T_0$, 
then $\g\in H^s((\re\setminus[-T-1,T+1])\times M)$ for $s<s'$. 
\end{lem}

Intuitively, the claim is straightforward from the above observation. 
The proof is elementary but slightly involved, and it is given in Appendix~A.

\subsection{Incoming observable}
\label{subsec-incoming-observable}
In this subsection, we construct an operator $B$ which is microlocally supported in the incoming region, 
and monotone decreasing (non increasing) along the null geodesics in the support. 

For $T\geq T_0$ in Lemma~\ref{lem-microlocalize}, we set 
\[
\z_1^\pm(t)=\i_1((\mp t/T)+1)
\]
so that $\z_1^\pm (t)=1$ if $\pm t\geq 2T$ and $\z_1^\pm(t)=0$ if $\pm t\leq T$. 

For $0<\d\ll 1$, we set
\[
\l(t)=2\d-\d\abs{t}^{-\n},
\]
for $|t|\geq 1$, where $0<\n<\m$. Note $\l(t)\geq \d$ and $\l(t)\to 2\d$ as 
$|t|\to\infty$, monotonically in $(-\infty,-1]$, 
and in $[1,\infty)$, respectively. We then set 
\[
\z_2^\pm(t,\t)= \i_2((\pm\t-1)/\l(t)). 
\]

We now set
\[
B=B(\d,T)=\sum_{\pm}\Op_h(|t|^\c\z_1^\mp \z_2^\pm) \i_2(\l(t)^{-1}(h^2Q_0-1)), \quad h>0. 
\]
We note $\i_2(\l(t)^{-1}(h^2Q_0-1))$ is an $h$-pseudodifferential operator with the principal symbol 
\[
\z_3(t,x,\x)=\i_2((q_0(x,\x)-1)/\l(t)).
\]

\begin{rem}
Crucial property of $B$ with this definition is that $B$ commutes with $Q_0$. 
This property does not hold if we define $B$ by straightforward quantization of 
the principal symbol 
\[
b_0(t,x,\t,\x)=\sum_{\pm}|t|^\c\z_1^\mp\z_2^\pm \z_3.
\] 
\end{rem}

\begin{lem}\label{lem-incoming-symbol}
Let $B=B(\d,T)$ as above. Then $i[B,P]$ has a symbol 
$f(t,x,\t,\x)\in S(\jap{t}^{\c-1},g_0)$ as an $h$-pseudodifferential operator, where $g_0=dt^2+dx^2+d\t^2+d\x^2$. 
The principal symbol of $i[B,P]$ is $-h^{-1}\{\t^2+q,b_0\}$, and it satisfies
\begin{align}\label{incpr0}
-\{\t^2+q,b_0\}\geq c_0|t|^{-1}b_0
\end{align}
with some constant $c_0>0$ if $T>0$ is sufficiently large.  
Moreover, $f$ is supported in $\supp[b_0]$ modulo the $O(h^\infty)$ terms. 
\end{lem}

\begin{proof} We note
\begin{align*}
i[B,P]=&i[B,D_t^2+Q]=ih^{-2}[B,(hD_t)^2+\Op_h(q)]\\
=&ih^{-2}[\Op_h(|t|^\c\z_1^\mp\z_2^\pm),(hD_t)^2+\Op_h(q)]\i_2(\l(t)^{-1}(h^2Q_0-1))\\
&+ih^{-2}\Op_h(|t|^\c\z_1^\mp\z_2^\pm)[\i_2(\l(t)^{-1}(h^2Q_0-1)),(hD_t)^2+\Op_h(q)]
\end{align*}
since $B$ commutes with $Q_0$. We first compute 
$i[\Op_h(|t|^\c\z_1^\mp\z_2^\pm),(hD_t)^2+\Op_h(q)]$
as an $h$-pseudodifferential operator. Its principal symbol is given by
\[
-h\{\t^2+q(t,x,\t,\x),|t|^\c\z_1^\mp(t)\z_2^\pm(t,\t)\}.
\]
It is easy to see
\[
-\{\t^2+q, |t|^\c\}= -(2\t+\pa_\t q) \c\, \mathrm{sgn}(t)|t|^{\c-1}
=2\c  |t|^{\c-1}(|\t|+O(\jap{t}^{-1-\m}))
\]
on the support of $\z_1^\mp(t)\z_2^\pm(t,\t)$, and this part gives us the positivity 
of the commutator. It remains to show the contribution from other terms are 
non negative. 

As well as the above computation, we have 
\begin{align*}
-\{\t^2+q,\z_1^\mp(t)\} &= \mp (2\t+\pa_\t q) T^{-1}\chi_1'((\pm t/T)+1)\\
&=2T^{-1}|\chi_1'((\pm t/T)+1)| (|\t|+O(\jap{t}^{-1-\m}))
\end{align*}
on the support of $\z_1^\mp(t)\z_2^\pm(t,\t)$, which is nonnegative. 
Then, Using $\l'(t)=\mathrm{sgn}(t)\d\n|t|^{-1-\n}$, we compute
\begin{align*}
&-\{\t^2+q,\z_2^\pm(t,\t)\} = -(2\t+\pa_\t q) \pa_t\z_2^\pm+\pa_t q\pa_\t\z_2^\pm\\
&\qquad =(2\t+\pa_\t q)\tfrac{(\pm \t -1)\l'(t)}{\l(t)^2}\i_2'\bigpare{\tfrac{\pm\t-1}{\l(t)}}
\pm \pa_t q\l^{-1}(t)\i_2'\bigpare{\tfrac{\pm\t-1}{\l(t)}} \\
&\qquad = \l^{-1}(t)\i_2'\bigpare{\tfrac{\pm\t-1}{\l(t)}}
\bigpare{(2\t+\pa_\t q)\mathrm{sgn}(t)\d\n |t|^{-1-\n} \bigpare{\tfrac{\pm \t -1}{\l(t)}}\pm\pa_t q(t,x,\x)}\\
&\qquad = \l^{-1}(t)\bigabs{\i_2'\bigpare{\tfrac{\pm\t-1}{\l(t)}}}
\bigpare{2\d\n|\t| |t|^{-1-\n} \bigabs{\tfrac{\pm \t -1}{\l(t)}}+O(\jap{t}^{-1-\m})}
\end{align*}
on the support of $\z_1^\pm(t)\z_2^\pm(t,\t)\z_3(t,x,\x)$, since $s\i_2'(s)\leq 0$. 
Since $\n<\m$ and $|\t|\sim 1$ on the support, 
we learn the right hand side is nonnegative on the support provided $T$ is chosen sufficiently large.  
Combining these, we have 
\begin{align}\label{incpr1}
-\{\t^2+q,|t|^\c\z_1^\pm\z_2^\pm\}\geq 2\c  |t|^{\c-1}\z_1^\pm(t)\z_2^\pm
\end{align}
on the support of $\z_1^\pm(t)\z_2^\pm(t,\t)\z_3(t,x,\x)$, provided $T$ is chosen sufficiently large. 

We then compute 
\[
i[\i_2\bigpare{\tfrac{h^2 Q_0-1}{\l(t)}}, D_t^2]
=-D_t[\pa_t,\i_2\bigpare{\tfrac{h^2 Q_0-1}{\l(t)}}]-[\pa_t,\i_2\bigpare{\tfrac{h^2 Q_0-1}{\l(t)}}]D_t 
\]
using the functional calculus. We have 
\begin{align*}
[\pa_t,\i_2\bigpare{\tfrac{h^2 Q_0-1}{\l(t)}}]&=-\frac{\l'(t)}{\l(t)^2}(h^2 Q_0-1)\i_2'\bigpare{\tfrac{h^2 Q_0-1}{\l(t)}}\\
&= - \frac{\mathrm{sgn}(t)\d\n|t|^{-1-\n}}{\l(t)^2}(h^2 Q_0-1)\i_2'\bigpare{\tfrac{h^2 Q_0-1}{\l(t)}}.
\end{align*}
Now we consider this as an $h$-pseudodifferential operator with the symbol in $S(h^{-1}\jap{t}^{-1-\n}\jap{\t},g_0)$. In particular, we note the principal symbol of 
$-i[\i_2(\l(t)^{-1}(h^2 Q_0-1)), D_t^2]$ is given by 
\[
h^{-1}\z_4(t,x,\t,\x)= \frac{2\t\,\mathrm{sgn}(t)\d\n|t|^{-1-\n}}{h\l(t)}
\frac{q_0-1}{\l(t)}\i_2'\Bigpare{\frac{q_0-1}{\l(t)}}.
\]
Noting that on the support of $\z_1^\mp\z_2^\pm$, $\t\,\mathrm{sgn}(t)<0$, and 
that $s\i_2'(s)\leq 0$, we learn $\z_1^\mp\z_2^\pm\z_4\geq 0$. 

Now the principal symbol of $ih^{-1}[\i_2(\l(t)^{-1}(h^2 Q_0-1)), \Op_h(q)]$ is given by 
\begin{align}
-\bigbra{q,\i_2\bigpare{\tfrac{q_0-1}{\l(t)}}}
&=\frac{\pa_\x q_0\cdot\pa_x q-\pa_x q_0\cdot\pa_\x q}{\l(t)}\i_2'\bigpare{\tfrac{q_0-1}{\l(t)}}\nonumber \\
&\qquad -\frac{\l'(t)(q_0-1)}{\l(t)^2}\i_2'\bigpare{\tfrac{q_0-1}{\l(t)}}\pa_\t q\nonumber\\
&=\bigabs{\i_2'\bigpare{\tfrac{q_0-1}{\l(t)}}}\cdot  O(\jap{(\t,\x)}^3|t|^{-1-\m}).\label{incpr2}
\end{align}
Thus the principal symbol of $ih^{-1}[\i_2(\l(t)^{-1}(h^2 Q_0-1)), (hD_t)^2+\Op_h(q)]$ is given by 
\begin{align}\label{incpr3}
\biggpare{\frac{2|\t|\d\n}{\l(t)}
\Bigabs{\frac{q_0-1}{\l(t)}}|t|^{-1-\n} +O(\jap{(\t,\x)}^3|t|^{-1-\m})}\Bigabs{\i_2'\Bigpare{\frac{q_0-1}{\l(t)}}}
\end{align}
on the support o $\z_1^\mp\z_2^\pm$, and it is non negative provided $T$ is chosen sufficiently large. 
Combining $(\ref{incpr1}), (\ref{incpr2})$ and $(\ref{incpr3})$, we conclude $(\ref{incpr0})$. 
\end{proof}

\subsection{Incoming regularity} \label{subsec-incoming-regularity}

In this subsection, we follow the argument of \cite{NT22} \S\S3.2, and show the 
$H^N$-regularity of $\g\in \Ker(P^*-z)$ in the incoming region. We only sketch the argument, 
and refer \cite{NT22} \S\S3.2 for the detail. 

Let $\d<\tilde\d$, $\tilde T<T$, and set $B=B(\d,T)$ and 
$\tilde B=B(\tilde\d,\tilde T)$. 
Then by the sharp G{\aa}rding inequality and Lemma~\ref{lem-incoming-symbol},
we learn there are $c,c'>0$ such that
\begin{equation}\label{eq-commutator-estimate-1}
i[B^*B,P]\geq \frac{c}{h} B^*\jap{t}^{-1}B - c'\tilde B^* \jap{t}^{-1}\tilde B - E^*E
\end{equation}
where $\norm{E}=O(h^\infty)$ as $h\to 0$. 

We set 
\[
\d_0<\d_1<\cdots <\d_\infty\ll 1, \quad T_0>T_1>T_2>\cdots>T_\infty \gg 0
\]
and let $B_j=B(\d_j,T_j)$. As in \cite{NT22} we can naturally show that 
\[
i[B_j^* B_j,P] \geq \frac{c_j}{h}B_j^*\jap{t}^{-1}B_j -c_j'B_{j+1}^*\jap{t}^{-1}B_{j+1}-E_j^*E_j
\]
for each $j$ with some $c_j,c_j'>0$ and $\norm{E_j}=O(h^\infty)$. Then this implies 
\begin{multline} \label{eq-key-estimate-1}
\frac{c_j}{2h}\norm{\jap{t}^{-1/2}B_j\f}^2+2(\Im z)\norm{B_j\f}^2  \\
\leq \frac{2h}{c_j}\norm{\jap{t}^{1/2}B_j (P-z)\f}^2 
+c_j'\norm{\jap{t}^{-1/2}B_{j+1}\f}^2 +\norm{E_j\f}^2.
\end{multline}

If we suppose $(P-z)\g=0$ where $\g\in L^2(X)$ and $\Im\, z>0$, then we use the standard 
bootstrap argument to show $\norm{B_0\g}=O(h^N)$ with any $N$ as $h\to 0$. 
See the proof of Lemma~3.3 in \cite{NT22} for the detail. 

\begin{lem}\label{lem-incoming-estimate}
Suppose Assumptions~\ref{ass-main} and \ref{ass-time-like}, $(P-z)\g=0$ with $\Im\, z>0$, 
and let $B_0$ as above. 
Then $\norm{B_0\g}_{L^2}=O(h^N)$ with any $N$ as $h\to 0$. 
\end{lem}

Now this implies that $\g$ is smooth in the incoming area. We set 
$\z_5^\pm(\t) = \i_1(1\mp\t)$ so that 
\[
\z_5^\pm(\t)=\begin{cases} 1 \quad&\text{if } \pm\t\geq1, \\ 0 \quad &\text{if }\pm\t\leq 0.\end{cases}
\]
We then set
\[
\Pi^\pm =\Op(\z_1^-(t)\z_5^\mp(\t)+\z_1^+\z_5^\pm(\t)),
\]
with $T=T_0$ so that $\Pi^-$ is a projection to microlocally 
incoming area (with $|t|>T$), and $\Pi^+$ 
is the projection to outgoing area. Here we use the usual (non semiclassical) quantization 
by $\Op(\cdot)$, i.e., the quantization with $h=1$. 
Recalling the symbol of $b_0$, we learn that Lemma~\ref{lem-incoming-estimate} implies 
the regularity of $\Pi^-\g$ by Lemma~\ref{lem-microlocalize}. More precisely, we have the 
following incoming regularity: 

\begin{lem}\label{lem-incoming-regularity}
Under the same setting as above, $\jap{t}^\c\Pi^-\g\in H^N(X)$ with any $N$. In particular, 
$\Pi^-\g\in C^\infty(X)$. 
\end{lem}

\subsection{Overall smoothness}
\label{subsec-overall-smoothness}

Combining the result of the previous subsection with the propagation of singularities theorem and Assumption~\ref{ass-time-like}, we learn $\g$ is smooth on $X$. At first we need a 
simple lemma:

\begin{lem}\label{lem-nontrapping-velocity-bound}
Suppose Assumption~\ref{ass-main} and  let $0<c_1<1$. Then there is $T_1>0$ such that 
if $p(t,x,\t,\x)=0$ and $|t|\geq T_1$, then $|2\t-\pa_\t p(t,x,\t,\x)|\leq c_1 |\t|$. 
In particular, $\t$ has the same sign as $\pa_\t p(t,x,\t,\x)$ if $|t|\geq T_1$. 
\end{lem}

\begin{proof}
It suffices to show $|2\t-\pa_\t p(t,x,\t,\x)|\leq c_1|\t|$ if $p(t,x,\t,\x)=0$. In fact, we then have
\[
\t(2-c_1{\t}/{|\t|})\leq \pa_\t p(t,x,\t,\x) \leq \t(2+c_1{\t}/{|\t|}).
\]
Since 
$2\pm c_1\t/|\t|\geq 2-c_1>0$, 
this implies $\t$ has the same sign as $\pa_\t p(t,x,\t,\x)$. 

Using the condition $p(t,x,\t,\x)=0$, we have 
$\t^2 -q_0(x,\x) =q(t,x,\t,\x)$, 
and hence, recalling the ellipticity of $q_0(x,\x)$, we learn 
\[
q_0(x,\x) = \t^2 -|q(t,x,\t,\x)| \leq \t^2 +C \jap{t}^{-1-\m}(\t^2+q_0(x,\x)). 
\]
This implies $q_0(x,\x)\leq 2\t^2$ if $|t|\gg 0$. Now we have
\[
|\pa_\t q(t,x,\t,\x)|\leq C \jap{t}^{-1-\m}(|\t|+|\x|)
\leq C'\jap{t}^{-1-\m}|\t|
\]
and this implies $|2\t-\pa_\t p(t,x,\t,\x)|\leq c_1 |\t|$, provided $|t|\gg 0$.
\end{proof}

Let $(t(s),x(s),\t(s),\x(s)))$ be a null geodesics. We recall the velocity is given by 
$(\dot t(s),\dot x(s))=(\pa_\t p,\pa_\x p)$. By Assumption~\ref{ass-time-like}, 
there are $s_b\ll 0$ such that either $t(s_b)<-\max(T_0,T_1)$ and $\dot t(s_b)>0$, 
or $t(s_b)>\max(T_0,T_1)$ and $\dot t(s_b)<0$. These imply, by 
Lemma~\ref{lem-nontrapping-velocity-bound},   $|t(s_b)|>T_0$ and 
$t(s_b)\t(s_b)<0$, i.e., incoming in the sense of the previous subsection. Hence, 
by Lemma~\ref{lem-incoming-regularity}, $(t(s_b),x(s_b),\t(s_b),\x(s_b)))\notin \WF(\g)$. 
Now we use the propagation of singularities theorem (\cite{Ho} Theorem~23.2.9):, and we learn that 
\begin{align*}
(t(s),x(s),\t(s),\x(s)))\notin \WF(\g)
\end{align*}
for all $s$. Thus we have proved: 

\begin{lem}\label{lem-global-regularity}
Let $\g\in L^2(X)$ such that $(P-z)\g=0$ with $\Im\, z>0$, then $\g\in C^\infty(X)$. 
\end{lem}

\subsection{Outgoing observable and the regularity}
\label{subsec-outgoing-observable-and-the-regularity}
While we now have the overall smoothness of $\g$, we actually need Sobolev estimates. 
We already have the Sobolev estimate for the incoming region, and it remains to show 
the Sobolev estimate for the outgoing region. We employ observables which are 
similar to those used in the incoming estimate, but somewhat different. 

We redefine the function $\l(t)$ as follows:
\[
\l(t)=\d+\d |t|^{-\n}
\]
for $t>1$, where $0<\d\ll 1$ and $0<\n<\m$ as well as in the incoming case. 
We note $\d\leq \l(t)\leq 2\d$, and $\l(t)\to \d$ as $|t|\to\infty$ monotonically in 
$t\in (1,\infty)$ and $(-\infty,-1)$, respectively. 
Then $\z_1^\pm$ and $\z_2^\pm$ are defined by the same 
expression but with the above $\l(t)$, i.e., 
\[
\z_1^\pm(t)=\i_1((\mp t/T)+1), \quad 
\z_2^\pm(t,\t)=\i_2((\pm\t-1)/\l(t)).
\]
We now set
\[
B=B(\d,T)=\sum_\pm \Op_h(|t|^{-\c}\z_1^\pm\z_2^\pm)\i_2(\l(t)^{-1}(h^2Q_0-1)), 
\]
where $h>0$ is the small semiclassical parameter as before, and $T\gg 0$. 
We write the principal symbol by $b_0(t,x,\t,\x)$, i.e.,
\[
b_0(t,x,\t,\x)=\sum_{\pm} |t|^{-\c}\z_1^\pm(t)\z_2^\pm(t,\t) \z_3(t,x,\x),
\]
where $\z_3(t,x,\x)=\i_2((q_0(x,\x)-1)/\l(t))$. 

We have a symbol estimate for this operator $B$, anaogously to 
Lemma~\ref{lem-incoming-symbol}, but slightly different. 

\begin{lem}\label{lem-outgoing-symbol}
Let $B=B(\d,R)$ as above. Then $i[B,P]$ has a symbol 
$f(t,x,\t,\x)\in S(\jap{t}^{-\c-1},g_0)$ as a $h$-pseudodifferential operator.
The principal symbol of $i[B,P]$ is $-h^{-1}\{\t^2+q,b_0\}$, and it satisfies
\[
-\{\t^2+q,b_0\}\geq c_0|t|^{-1}b_0-\tilde f
\]
with some constant $c_0>0$ if $T>0$ is sufficiently large, where $\tilde f(t,x,\t,\x)\in S(1,g_0)$ is supported 
in $\{(t,x,\t,\x)\,|\, T\leq t\leq 2T\}\cap \supp[b_0]$. 
Moreover, $\tilde f$ is supported in $\supp[b_0]$ modulo the $O(h^\infty)$ terms. 
\end{lem}

\begin{proof}[Sketch of Proof]
The proof is almost the same as that of Lemma~\ref{lem-outgoing-symbol}, but there are 
several differences, and we sketch them here. 

At first, the Poisson bracket $-\{\t^2+q,\z_1^\pm\}$ is not necessarily nonnegative 
on $\supp[\z_1^\pm\z_2^\pm]$. In fact, it is nonpositive, and hence we set 
\[
\tilde f(t,x,\x)=\{\t^2+q,\z_1^\pm\}\z_2^\pm\z_3
\]
to compensate it. 

The other computations are almost the same except for the fact 
$\l'(t)=-\mathrm{sgn}(t)\d\n|t|^{-\n-1}$, 
i.e., the sign is opposite from the incoming case. However it is consistent since 
$t\t>0$ on $\supp[\z_1^\pm\z_2^\pm]$, and almost the same computations are 
carried out with changes of several signs. We omit the detail. 
\end{proof}

We then proceed to show regularity in the outgoing region. As well as in the incoming case, 
in particular as \eqref{eq-commutator-estimate-1}, we let
$\d<\tilde\d$, $\tilde T<T$, and set $B=B(\d,T)$ and 
$\tilde B=B(\tilde\d,\tilde T)$. Then by Lemma~\ref{lem-outgoing-symbol} 
and the sharp G{\aa}rding inequality, we learn there are $c_0,c'>0$ such that
\[
i[B^*B,P]\geq \frac{c_0}{h} B^*\jap{t}^{-1}B - c'\tilde B^* \jap{t}^{-1}\tilde B - E^*E 
-F
\]
where $F=B^*\Op_h(\tilde f)+\Op_h(\tilde f)^* B$ and  $\norm{E}=O(h^\infty)$ as $h\to 0$. 
As well as \eqref{eq-key-estimate-1}, this implies, 
\begin{multline*} 
\frac{c_0}{2h}\norm{\jap{t}^{-1/2}B\f}^2+2(\Im z)\norm{B\f}^2  \\
\leq \frac{2h}{c_0}\norm{\jap{t}^{1/2}B (P-z)\f}^2 
+c'\norm{\jap{t}^{-1/2}\tilde B\f}^2 +\norm{E\f}^2+\jap{\f,F\f}
\end{multline*}
for $\f\in C_0^\infty(X)$. 
This estimate is easily extended to $\g\in L^2$ with $(P-z)\g=0$, and we learn 
\[
\frac{c_0}{2h}\norm{\jap{t}^{-1/2}B\g}^2+2(\Im z)\norm{B\g}^2
\leq c'\norm{\jap{t}^{-1/2}\tilde B\g}^2 +\norm{E\g}^2 +\jap{\g,F\g}.
\]
We observe $\norm{E\g}=O(h^N)$ with any $N$ by the construction of $E$.
We now recall the support of the symbol of $F$ is contained in $\{T\leq t\leq 2T\}\cap \supp[b_0]$, 
and in particular, compactly supported in $X$. 
By this and the global regularity, i.e., Lemma~\ref{lem-global-regularity}, we learn $\jap{\g,F\g}=O(h^N)$ 
with any $N$. 
Thus we can use the same iteration procedure as in Subsection~\ref{subsec-incoming-regularity} 
to conclude $\norm{B_0\g}_{L^2}=O(h^N)$ with any $N$ as $h\to 0$ as well as 
Lemma~\ref{lem-incoming-estimate}. 
Now we have the following outgoing regularity result as well as the incoming case, 
i.e., Lemma~\ref{lem-incoming-regularity}: 

\begin{lem}\label{lem-outgoing-regularity}
Suppose $\g\in L^2(X)$ and $(P-z)\g=0$ where $\Im\, z>0$. 
Then $\jap{t}^{-\c}\Pi^+\g\in H^N(X)$ with any $N$, provided $T$ is sufficiently large. 
\end{lem}

\subsection{Proof of Theorem~\ref{thm-main}}
\label{subsec-proof-of-theorem-main}

Let $\g\in L^2(X)$ and $(P-z)\g=0$ with $\Im\, z>0$. 
Then, combining 
Lemmas~\ref{lem-incoming-regularity}, \ref{lem-global-regularity} and
\ref{lem-outgoing-regularity}, we learn that $\jap{t}^{-\c}\g\in H^N(X)$ with
any $N$. Then, by Lemma~\ref{lem-first-reduction}, we conclude $\g=0$. 
Thus we have proved $\mathrm{Ker}(P^*-z)=\{0\}$. 
By similar arguments, we can also show $\mathrm{Ker}(P^*-z)=\{0\}$
when $\Im\, z<0$. These implies the essential self-adjointness of $P$ on $C_0^\infty(X)$ 
(see, e.g.,   \cite{RS2} Theorem~X.1). \qed

\appendix
\section{Proof of Lemma~\ref{lem-microlocalize}}
\label{app-microlocalize-proof}
Let $\y(t,x,\t,\x)=\i_2((\t^2+q_0(x,\x)-2)/\c)$ with $\c=\d/4$. It suffices to show 
$\norm{\Op_h(\y)\g}_{L^2(\tilde X)}=O(h^{s'})$ as $h\to 0$,  where 
$\tilde X=(\re\setminus[-T-1,T+1])\times M$ with $T>T_0$. 
We choose $T_0$ so that 
\[
|q(t,x,\t,\x)|\leq \a (\t^2+q_0(x,\x))\quad\text{for }|t|\geq T_0
\]
with $\a=\d/(4+\d)$ (see Assumption~\ref{ass-main}). 
By the assumption, we know 
$\norm{\Op_h(\y) \Op_h(a(\d,T))\g}_{L^2(\tilde X)}=O(h^{s'})$, 
and hence it remains to show 
\begin{equation}\label{eq-lem-microlocal-key}
\norm{\Op_h(\y)\Op_h(1-a(\d,T))\g}_{L^2(\tilde X)}=O(h^{s'}),
\end{equation}
where $T\geq T_0$. We note the symbol of $\Op_h(\y)\Op_h(1-a(\d,T))$ is essentially 
supported in the support of $\y(1-a(\d,T))$, and if $P$ is elliptic on this support, 
the property \eqref{eq-lem-microlocal-key} follows from the equation $(P-z)\g=0$ by the standard elliptic estimate
(with any $s'>0$). 
Thus, in order to prove \eqref{eq-lem-microlocal-key}, it is sufficient to show 
$p(t,x,\t,\x)\geq \d>0$ on the support of  $\y(1-a(\d,T))$. 

By the construction, we have 
\begin{equation}\label{eq-app-est-1}
|\t^2+q_0-2|\leq 2\c \quad \text{on }\supp[\y].
\end{equation}
We also have 
\begin{equation}\label{eq-app-est-2}
|q|\leq \a(\t^2+q_0)\leq \a(2+2\c) \quad \text{on }\supp[\y]\cap\{|t|\geq T_0\}.
\end{equation}
using the choice of $T_0$ and \eqref{eq-app-est-1}. 

On the other hand, we have  
\begin{equation}\label{eq-app-est-3}
|\t^2-1|\geq \d \text{\ \ or\ \ } |q_0-1|\geq \d 
\quad \text{on }\supp[1-a]\cap\{|t|\geq T\}.
\end{equation}
If $|\t^2-1|\geq \d$, then we have 
\begin{align*}
|p| &= |\t^2-q_0+q|\geq |\t^2-q_0|-|q|\\
&\geq |\t^2 -(2-\t^2)|-2\c -\a(2+2\c) \\
&=2|\t^2-1|-(2\c+\a(2+2\c)) =2|\t^2-1|-\d
\geq \d,
\end{align*}
where we have used \eqref{eq-app-est-2} and \eqref{eq-app-est-3} in the second inequality. 
We note $2\c+\a(2+2\c)=\d$ by our choice of constants $\a,\c$. 
Similarly, if $|q_0-1|\geq\d$, then we have
\begin{align*}
|p| &= |\t^2-q_0+q|\geq |\t^2-q_0|-|q|\\
&\geq |(2-q_0)-q_0|-2\c -\a(2+2\c) \\
&=2|q_0-1|-(2\c+\a(2+2\c)) =2|q_0-1|-\d
\geq \d,
\end{align*}
using \eqref{eq-app-est-2} and \eqref{eq-app-est-3} again. 
These inequalities imply $P$ is elliptic on $\supp[\y(1-a(\d,T))]\cap\{|t|\geq T\}$, and 
this completes the proof of Lemma~\ref{lem-microlocalize}. 
\qed


\begin{thebibliography}{99}
\bibitem{DeSi1} Derezi\'nski, J., Siemssen, D.: Feynman propagators on static spacetimes, Rev. Math. Phys. \textbf{30}, (2018), 1850006.
\bibitem{DeSi2} Derezi\'nski, J., Siemssen, D.:  An evolution equation approach to the Klein-Gordon operator on curved spacetime, Pure Appl. Anal. \textbf{1}, 215--261, (2019).
\bibitem{DeSi3} Derezi\'nski, J., Siemssen, D.:  An Evolution Equation Approach to Linear Quantum Field Theory, preprint, arXiv:1912.10692, (2019).
\bibitem{DiSj} Dimassi, M., Sj\"ostrand, J.: Spectral Asymptotics in the Semi-Classical Limit. 
Cambridge Univ. Press, London Math. Soc. Lecture Note Series \textbf{268}, 1999. 
\bibitem{GeWr1} G\'erard, C., Wrochna, M.: The massive Feynman propagator on asymptotically Minkowski spacetimes, Amer. J. Math. \textbf{141}, (2019), 1501--1546.
\bibitem{GeWr2} G\'erard, C., Wrochna, M.: The massive Feynman propagator on asymptotically Minkowski spacetimes II, Int. Math. Res. Notices. \textbf{2020}, (2020), 6856--6870.
\bibitem{Ho} H\"ormander, L.: {\it Analysis of Linear Partial Differential Operators}, Vol.\ I-IV.  Springer Verlag, 1983--1985. 
\bibitem{NT21} Nakamura, S., Taira, K.: Essential self-adjointness of real principal type operators. Annales Henri Lebesgue, \textbf{4} (2021), 1035--1059.
\bibitem{NT22} Nakamura, S., Taira, K.: A remark on the essential self-adjointness for Klein-Gordon type operators. Preprint, arXiv: 2202.13499 (2022). 
\bibitem{RS2}Reed, M., Simon, B.: The Methods of Modern Mathematical Physics. 
Vol.~2. Fourier Analysis, Self-Adjointness. Academic Press 1975.
\bibitem{Va} Vasy, A.: Essential self-adjointness of the wave operator and the limiting absorption 
principle on Lorentzian scattering spaces. J. Spectr. Theory \textbf{10} (2020), no. 2, 439--461. 
\bibitem{Zw} Zworski, M.: Semiclassical Analysis. American Math. Soc., GSM \textbf{138}, 2012. 
\end{thebibliography}
\end{document}